\documentclass[fleqn,10pt]{wlscirep}

\usepackage[utf8]{inputenc}
\usepackage[T1]{fontenc}
\usepackage{amsmath,amsfonts,amssymb}
\usepackage{graphicx}
\usepackage{bm}
\usepackage{xcolor}
\usepackage{siunitx}
\usepackage{booktabs}
\usepackage{url}

\newcommand{\bfs}{BaFe$_2$S$_3$}
\newcommand{\bfse}{BaFe$_2$Se$_3$}
\newcommand{\revised}[1]{#1}

\title{\revised{Coexistence of superconductivity and magnetic order in BaFe$_2$S$_3$: evidence for orbital-selective delocalization}}

\author[1,2]{Y. Oubaid}
\author[1,3]{N. S. Dhami}
\author[2]{P. Fertey}
\author[4]{C. V. Colin}
\author[5]{C. Ritter}
\author[6]{A. Forget}
\author[7]{D. Bounoua}
\author[2]{J.-P. Rueff}
\author[2]{M. Verseils}
\author[1]{P. Foury-Leylekian}
\author[1,8,*]{V. Bal\'edent}

\affil[1]{Universit\'e Paris-Saclay, CNRS, Laboratoire de Physique des Solides, 91405 Orsay, France}
\affil[2]{Synchrotron SOLEIL, L'Orme des Merisiers, Saint Aubin BP 48, 91192 Gif-sur-Yvette, France}
\affil[3]{Institute of Physics, Bijenicka cesta 46, 10000 Zagreb, Croatia}
\affil[4]{Institut N\'eel, Universit\'e Grenoble Alpes \& CNRS, 38042 Grenoble, France}
\affil[5]{Institut Laue-Langevin, 71 avenue des Martyrs, 38042 Grenoble, France}
\affil[6]{Universit\'e Paris-Saclay, CEA, CNRS, SPEC, 91191 Gif-sur-Yvette, France}
\affil[7]{Laboratoire Leon Brillouin, CEA, CNRS, Universit\'e Paris-Saclay, CEA Saclay, 91191 Gif-sur-Yvette, France}
\affil[8]{Institut universitaire de France (IUF)}

\affil[*]{victor.baledent@universite-paris-saclay.fr}


\begin{abstract}
Understanding how superconductivity emerges from correlated electronic states is a central challenge in condensed matter physics. The quasi-one-dimensional spin-ladder compound \bfs\ serves as a unique platform to study this interplay, as it lacks the two-dimensional square lattice characteristic of most iron-based superconductors. Here, we combine high-pressure x-ray and neutron diffraction, x-ray emission spectroscopy, and electrical transport to map its structural, magnetic, and electronic evolution into the superconducting regime. We \revised{observe} a striking pressure-induced decoupling: while the long-range stripe-type antiferromagnetic order remains remarkably robust under pressure up to the onset of superconductivity, the local iron magnetic moment is suppressed by nearly 80\%. This dichotomy \revised{points to an orbital-selective mechanism}, where specific Fe $3d$ electrons remain localized to sustain magnetism while others delocalize to form the itinerant and superconducting state. \revised{Superconductivity emerges while long-range magnetic order is still present, revealing that the two coexist in \bfs, which we attribute to} the selective response of multiorbital electronic degrees of freedom.
\end{abstract}

\begin{document}

\flushbottom
\maketitle
\thispagestyle{empty}

\section*{Introduction}

Since the discovery of high-$T_c$ cuprates, the search for unconventional superconductors has expanded to iron-based compounds, heavy fermion systems, and nickelates~\cite{Scalapino2012,Keimer2015,Stewart2017,Stewart2011}. A widely shared paradigm is that superconductivity must be understood in close connection with neighboring electronic orders --- magnetism, charge ordering, or nematicity --- and that investigating how these orders evolve upon approaching the superconducting dome is essential for identifying the microscopic ingredients of Cooper pairing~\cite{Fernandes2014,Fradkin2015}.

This strategy has been particularly fruitful in quasi-two-dimensional iron-based superconductors such as FeSe and BaFe$_2$As$_2$, where superconductivity emerges from magnetically ordered parent phases under chemical substitution or applied pressure~\cite{Johnston2010,Stewart2011,Kordyuk2012}. However, the predominance of layered square-lattice geometries has constrained the exploration of how reduced dimensionality and strong correlations reshape this interplay.

The discovery of pressure-induced superconductivity in the quasi-one-dimensional spin-ladder compounds BaFe$_2$X$_3$ (X~=~S,~Se) opened a new frontier in iron-based superconductivity~\cite{Takahashi2015,Ying2017,yamauchi2015pressure}. These materials are the first iron-based superconductors lacking a two-dimensional square lattice, instead consisting of two-leg ladders formed by edge-sharing FeX$_4$ tetrahedra, with highly anisotropic transport and low-dimensional magnetic correlations~\cite{Lei2011,Nambu2012,Mourigal2015,Herbrych2018}.

Among these systems, \bfs\ is particularly intriguing. At ambient pressure, it is a Mott insulator displaying stripe-type antiferromagnetic order below $T_{\mathrm{N}} \approx 120$~K, with an ordered moment of about 1.3--1.5~$\mu_{\mathrm{B}}$ per Fe~\cite{Caron2012}. Applying pressure drives \bfs\ through an insulator-to-metal transition and into a superconducting state with $T_c \approx 24$~K~\cite{Takahashi2015}. The central question is whether superconductivity emerges from the suppression of magnetism or whether the two orders can intrinsically coexist in this quasi-one-dimensional setting.

The multiorbital character of the Fe~$3d$ shell and the reduced dimensionality of the ladder structure have led theoretical studies to predict that BaFe$_2$X$_3$ lies close to an orbital-selective Mott phase, in which some $3d$ orbitals are Mott localized while others remain metallic~\cite{Arita2015,Nambu2015,Herbrych2020,Craco2018}, with supporting spectroscopic evidence~\cite{Ootsuki2015}. Multiorbital Hubbard and density-matrix renormalization group calculations show that Hund's coupling and anisotropic hopping can stabilize a state where stripe magnetism coexists with partially itinerant carriers~\cite{Craco2018}. In the related compound \bfse, resonant inelastic x-ray scattering and neutron experiments suggest a reduction of the local moment under pressure without a proportional suppression of magnetic order~\cite{Zheng2022}, reinforcing the notion of orbital selectivity.

Pressure provides a clean tuning parameter to control the relative weight of localized and itinerant electronic components, by continuously modifying bandwidths and crystal-field splittings. To address whether superconductivity arises from orbital delocalization, magnetic quantum criticality, or the coexistence of distinct electronic degrees of freedom, it is crucial to disentangle local magnetic moments from long-range magnetic order and to correlate their evolution with structural changes under pressure.

Here we combine x-ray diffraction, neutron diffraction, x-ray emission spectroscopy (XES), and electrical resistivity measurements on \bfs\ under pressure from ambient conditions up to the superconducting regime. We \revised{observe} a clear separation between the pressure evolution of the local moment and that of the long-range ordered moment, \revised{which is consistent with} an orbital-selective mechanism underlying the coexistence of magnetism and superconductivity.

\section*{Results}

\subsection*{Low-pressure structural transition}

To understand the evolution of the electronic properties and the eventual emergence of superconductivity in \bfs, it is fundamental to first characterize the structural response to pressure, as subtle lattice anomalies often act as precursors to electronic or magnetic instabilities. At ambient pressure and low temperature, \bfs\ crystallizes in the orthorhombic $Pb2_1m$ space group, characterized by a specific tilting and distortion of the Fe$_2$S$_3$ ladders~\cite{Oubaid2026}. To track the evolution of this symmetry as a function of pressure, we performed single-crystal x-ray diffraction using diamond-anvil cells (DACs) down to 15~K on the CRISTAL beamline (Synchrotron SOLEIL). By using a helium transmitting medium, we maintained quasi-hydrostatic conditions, which are essential to avoid broadening of the Bragg reflections and to precisely identify the symmetry-breaking peaks. Experimental details and structural refinement data are provided in the Supplementary Information.

\begin{figure}[htbp]
\centering
\includegraphics[width=0.85\linewidth]{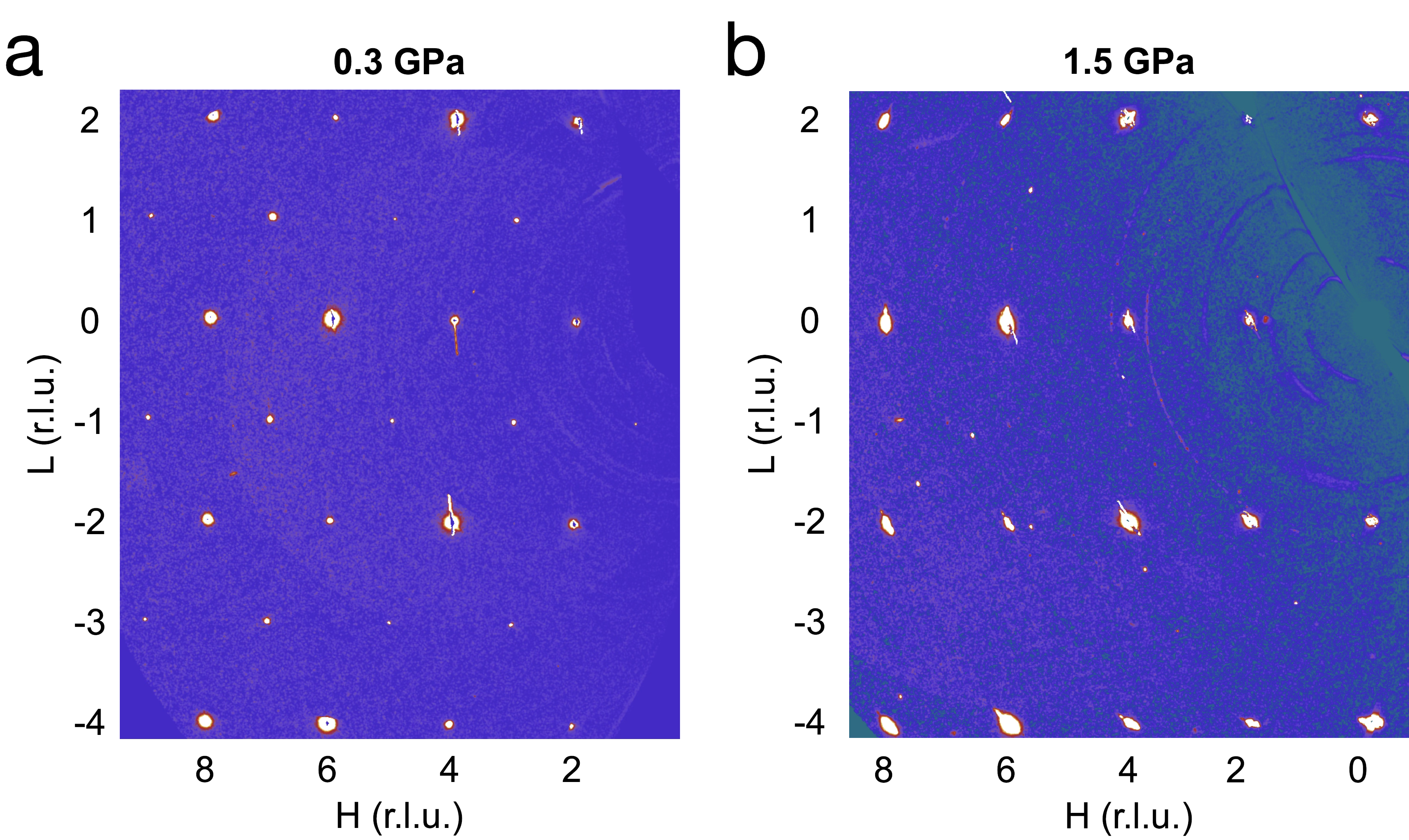}
\caption{\textbf{Structural transition under pressure.} Reconstruction of the reciprocal plane $(H0L)$ at 15~K under (a)~0.3~GPa and (b)~1.5~GPa. At 0.3~GPa, Bragg reflections breaking the extinction conditions of the $C$ centering ($h+k \neq 2n$) expected in the $Cmcm$ space group are present at odd $H$ values (for $K=0$), and are absent at 1.5~GPa.}
\label{Fig_HPXRD}
\end{figure}

Our data reveal a clear structural phase transition near 1~GPa. At low pressure (0.3~GPa, 15~K), we observe Bragg reflections characteristic of the $Pb2_1m$ symmetry. Specifically, the presence of reflections violating the $C$-centering extinction conditions ($h+k \neq 2n$) of the higher-symmetry $Cmcm$ group confirms the low symmetry of the ambient-pressure phase [Fig.~\ref{Fig_HPXRD}(a)]. Upon increasing pressure to 1.5~GPa, these specific reflections vanish completely [Fig.~\ref{Fig_HPXRD}(b)], signaling a transition to the $Cmcm$ space group. This transition corresponds to a straightening of the Fe-ladder framework and an increase in the overall symmetry of the Fe-S coordination environment. Crucially, our measurements confirm that the $Cmcm$ symmetry remains stable upon further compression, with no further structural transitions observed up to the highest investigated pressure of 16.5~GPa, well above the critical superconducting pressure (see Supplementary Information).

The significance of the initial transition at 1~GPa is twofold. First, the transition pressure ($\sim 1$~GPa) coincides remarkably well with the previously reported anomaly in the pressure dependence of the N\'eel temperature $T_{\mathrm{N}}$ and a localized jump in the ordered magnetic moment~\cite{Chi2016}. Second, the shift to $Cmcm$ symmetry implies a modification of the Fe-S-Fe bond angles and Fe-Fe distances. These structural parameters directly govern the magnetic exchange integrals ($J$) both along the legs and the rungs of the ladders. The transition to the $Cmcm$ phase thus provides a natural structural basis for the reinforcement of magnetic correlations observed in this pressure range, as the straightening of the Fe-S-Fe angle enhances superexchange interactions, before the eventual delocalization of electrons at much higher pressures.

\subsection*{Persistence of long-range magnetic order}

While the low-pressure structural transition sets the stage, the central enigma of \bfs\ lies in the fate of its magnetism as it approaches the superconducting dome. Previous studies were limited to low pressure~\cite{Caron2012,Chi2016}. To address the interplay between magnetism and superconductivity, it is imperative to probe the long-range magnetic order up to the highest possible pressures. The magnetic Bragg peaks directly track the \revised{long-range ordered component of the Fe moments}, whose spins generate the stripe-type antiferromagnetic (AFM) order. To achieve this, we performed powder neutron diffraction measurements up to 10~GPa using a Paris-Edinburgh (PE) press inserted in a cryostat, reaching base temperatures of 20~K at 2~GPa and 10~K at higher pressures. The pressure uncertainty for each data point is smaller than 0.3~GPa (see Supplementary Information).

\begin{figure}[htbp]
\centering
\includegraphics[width=0.85\linewidth]{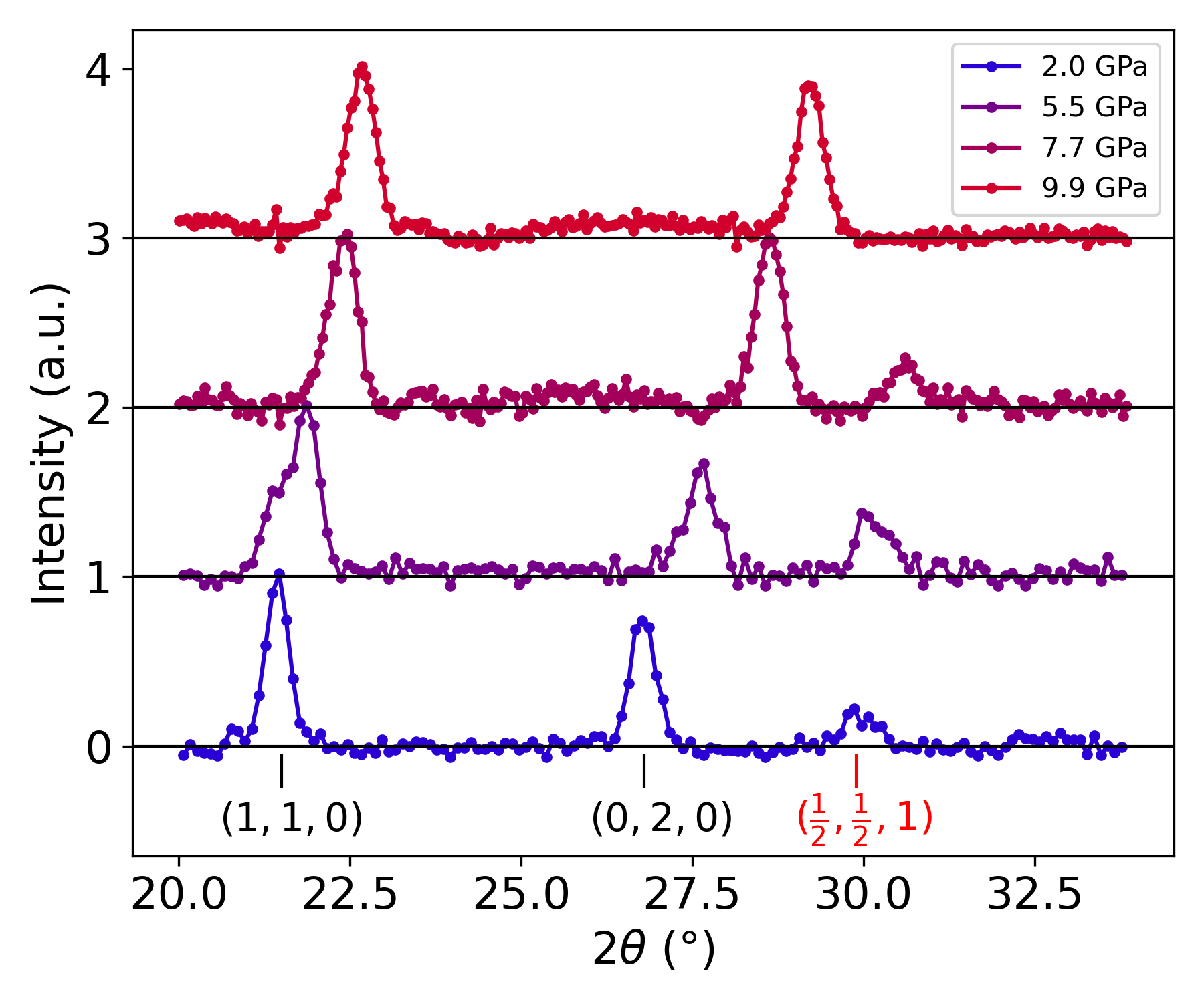}
\caption{\textbf{Neutron diffraction under pressure.} Powder neutron diffraction patterns of \bfs\ collected at 10~K (20~K for 2 and 5.5~GPa) under applied pressures ranging from 2 to 9.9~GPa. Only low-angle reflections are shown, labelled with their Miller indices $(hkl)$. Patterns collected at 2 and 5.5~GPa were measured on the D1B diffractometer ($\lambda = 2.52$~\AA), whereas those at 7.7 and 9.9~GPa were measured on D20 ($\lambda = 2.42$~\AA). All patterns are rescaled to a common wavelength of 2.52~\AA\ to facilitate comparison. Background has been subtracted and patterns are vertically offset for clarity.}
\label{Fig_HPNPD}
\end{figure}

Our results \revised{show} that the stripe-type AFM structure remains remarkably robust up to \revised{7.7~GPa} (Fig.~\ref{Fig_HPNPD}). Surprisingly, the magnetic reflections do not show significant broadening or a gradual decrease in intensity that would suggest simple melting of the magnetic order. Instead, the long-range ordered moment, $\mu_{\mathrm{ord}}$, remains nearly constant at $1.4 \pm 0.2~\mu_{\mathrm{B}}$ until it is abruptly suppressed between 7.7 and 9.9~GPa. This persistence suggests that the localized electrons responsible for the magnetic order are shielded from the initial effects of metallization reported in the literature~\cite{Takahashi2015}, maintaining a stable magnetic skeleton even as the system approaches superconductivity.

\subsection*{Pressure evolution of the local magnetic moment}

The stability of the long-range ordered moment ($\mu_{\mathrm{ord}}$) observed by neutrons appears paradoxical in light of the rapid crossover from insulating to metallic behavior reported in transport measurements~\cite{Takahashi2015}. To resolve this, we must distinguish between the long-range spatial correlation of spins and the magnitude of the local magnetic moment, which encompasses both the localized and the increasingly itinerant Fe $3d$ electrons. To probe this local spin state, we utilized Fe $K\beta$ x-ray emission spectroscopy (XES) under high pressure. Unlike neutron diffraction, which is sensitive only to the static, long-range ordered component, XES is an element-selective local probe that measures the local magnetic moment via the exchange interaction between the Fe $3p$ core hole and the entire $3d$ valence shell~\cite{Peng1994,Rueff1999,Rueff2001,Vanko2006a,Lafuerza2020}. Measurements were performed at the GALAXIES beamline (Synchrotron SOLEIL) using a membrane-driven DAC, tracking the total magnetic moment of the iron atoms --- encompassing both localized and itinerant contributions --- as the electronic bandwidth increases under compression. Further experimental details are given in the Supplementary Information.

As shown in Fig.~\ref{Fig_HPXES}, the XES spectra reveal a dramatic and continuous suppression of the local magnetic moment. The satellite peak ($K_{\beta'}$) intensity, \revised{which scales with the local spin moment~\cite{Vanko2006a}}, decreases steadily from ambient pressure. By 5~GPa, the local moment is already reduced to half its ambient-pressure value, and at 8~GPa it falls to 30\% of its initial value, despite a robust ordered moment of $1.4~\mu_{\mathrm{B}}$ being maintained. Above 10~GPa, the local moment reaches a plateau at approximately 20\% of its ambient-pressure value. This striking decoupling --- the collapse of the total local moment probed by XES against the constant ordered moment --- \revised{is naturally accounted for by} an orbital-selective mechanism. \revised{In this picture,} while some orbitals delocalize and magnetically vanish to contribute to the itinerant Fermi sea, a specific subset remains robust and localized, sustaining the long-range magnetic skeleton.

\begin{figure*}[htbp]
\centering
\includegraphics[width=0.75\linewidth]{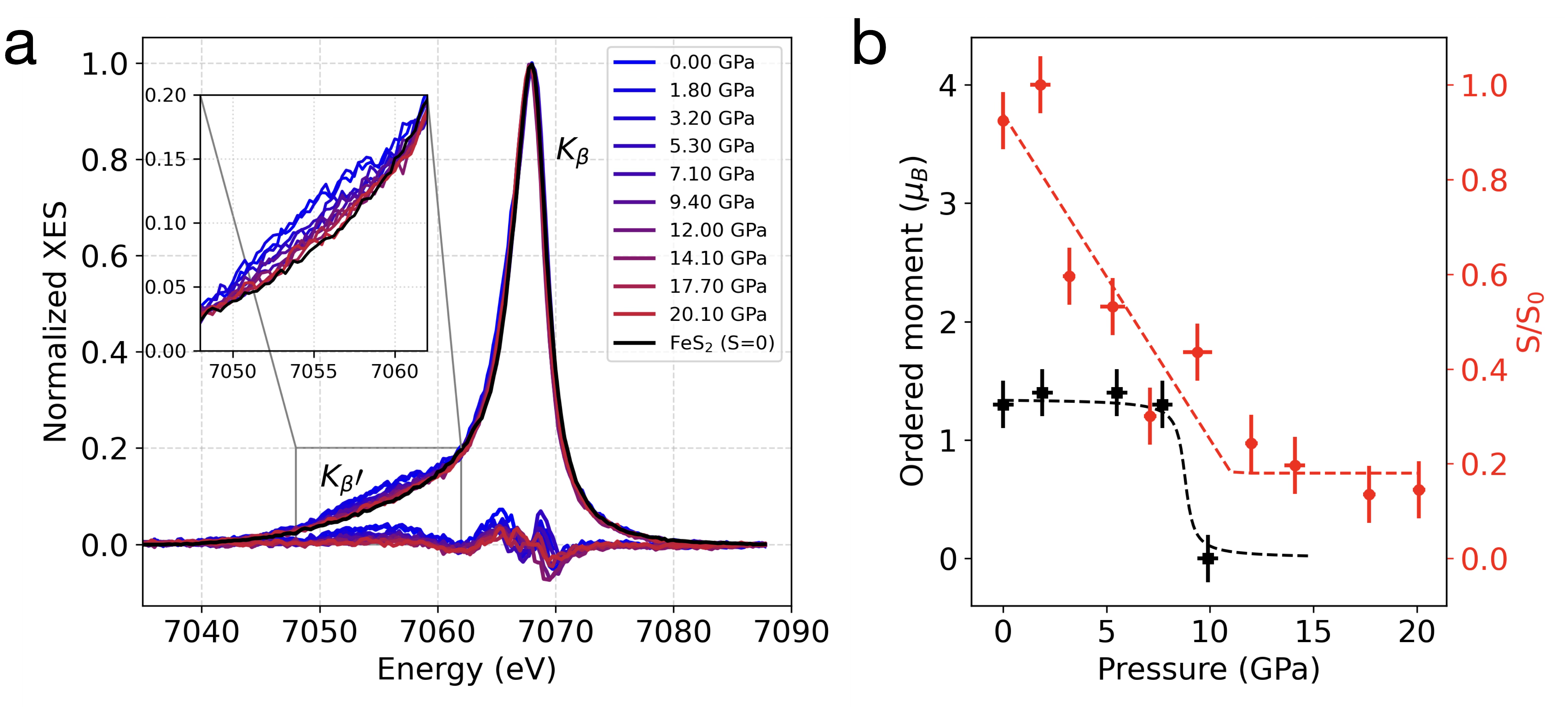}
\caption{\textbf{Decoupling of local and ordered magnetic moments.} (a)~X-ray emission spectra (XES) of the Fe $K_{\beta}$ line at 300~K for \bfs\ and FeS$_2$ ($S = 0$ reference) at various pressures. The inset shows a magnification of the $K_{\beta'}$ satellite region at $\sim$7055~eV. (b)~Pressure dependence of the Fe ordered magnetic moment from powder neutron diffraction (PND) below 20~K (black squares, left axis); error bars represent refinement uncertainty. Red circles (right axis) show the local spin $S$ at 300~K, extracted via integrated absolute difference (IAD) analysis and normalized to the ambient-pressure value $S_0$. Error bars are derived from photon statistics within the 7050--7060~eV window. Dashed lines are guides to the eye.}
\label{Fig_HPXES}
\end{figure*}

\subsection*{Transport and superconductivity}

\begin{figure*}[htbp]
\centering
\includegraphics[width=0.6\linewidth]{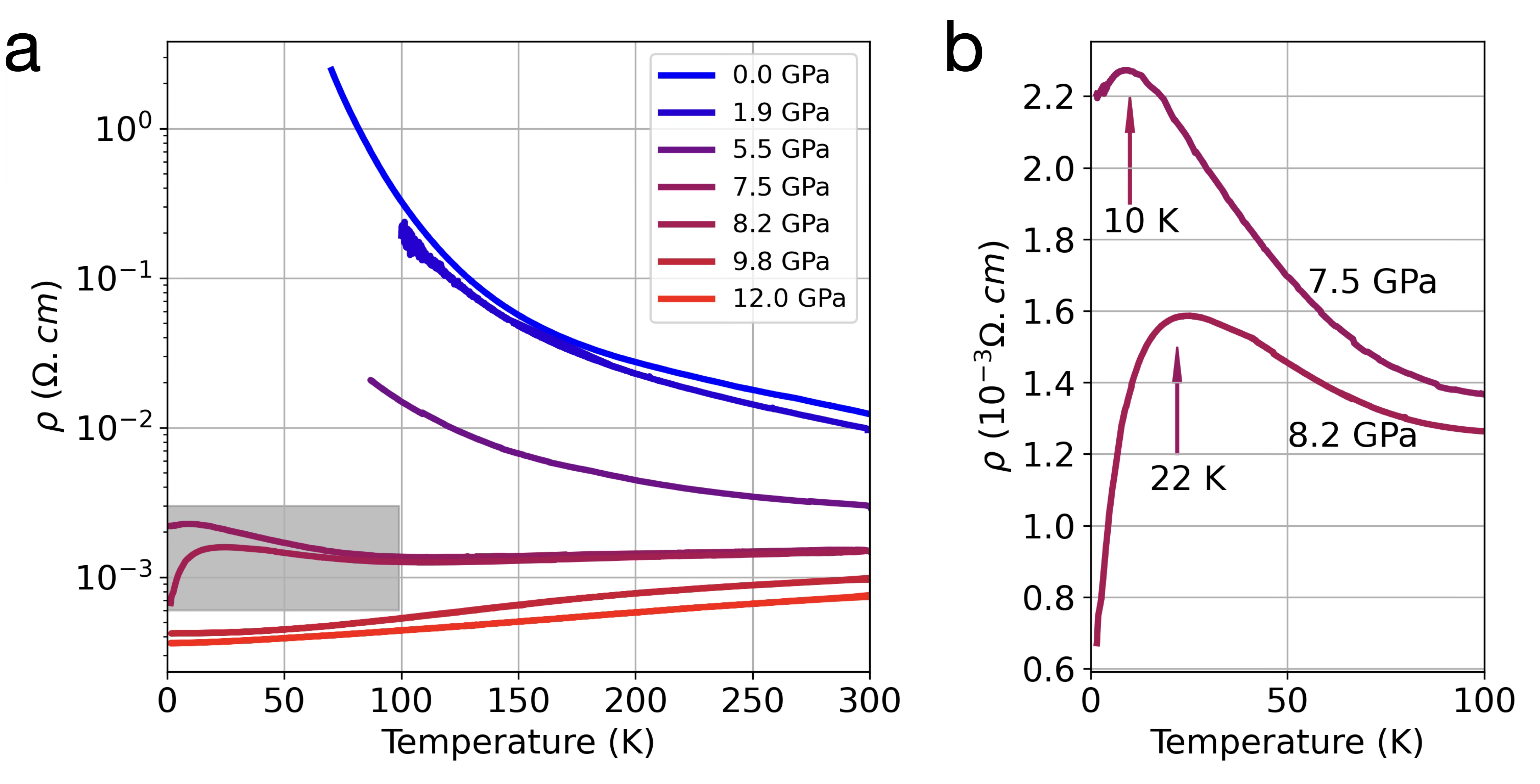}
\caption{\textbf{Electrical resistivity under pressure.} (a)~Temperature dependence of the resistivity of \bfs\ from ambient pressure to 12~GPa. The evolution from insulating-like to metallic transport is clearly observed. (b)~Enlarged view of the region at 7.5 and 8.2~GPa. Arrows indicate the onset of superconductivity at approximately 10~K and 22~K, respectively.}
\label{Fig_HPR}
\end{figure*}

To ensure the validity of the comparison between local probes and macroscopic properties, electrical transport measurements were performed on crystals from the same synthesis batch, using the standard four-probe DC technique in a DAC with Daphne 7373 as the pressure-transmitting medium. The results show a gradual suppression of the Mott insulating state, as evidenced by progressive deviation from activated behavior shown in Fig.~\ref{Fig_HPR}(a). Around 7~GPa, a high-temperature metallic response develops while a low-temperature semiconducting behavior persists, indicating incomplete metallization. Superconductivity emerges from this anomalous normal state with an onset $T_c \approx 10$~K at 7.5~GPa, strongly enhanced to approximately 22~K at 8.2~GPa. Similar behavior near superconductivity has been reported in earlier studies of BaFe$_2$X$_3$ (X = S, Se)~\cite{Takahashi2015,Ying2017} and is reminiscent of transport in organic Mott insulators~\cite{taniguchi2003superconductivity,vuletic2002coexistence}.

Notably, superconductivity \revised{is observed at 7.5 and 8.2~GPa and is absent at 5.5 and 9.8~GPa. Taking the midpoints between these pressures as bounds, the superconducting range extends from} $6.5 \pm 1$~GPa to $9.0 \pm 0.8$~GPa, \revised{and overlaps with the pressure range} in which neutron diffraction still detects long-range antiferromagnetic order. \revised{Superconductivity therefore sets in while long-range antiferromagnetic order is still present, showing that the two phases coexist.} At higher pressures (9.8 and 12~GPa), the low-temperature resistivity follows Fermi-liquid behavior ($\rho \propto AT^2$), with the quadratic coefficient $A$ increasing from $9.3 \times 10^{-6}$~m$\Omega$~cm~K$^{-2}$ at 9.8~GPa to $12.7 \times 10^{-6}$~m$\Omega$~cm~K$^{-2}$ at 12~GPa (see Supplementary Information).

\section*{Discussion}

Our results support a scenario in which orbital-selective correlations in \bfs\ give rise to two distinct electron populations among the Fe~$3d$ states --- one of localized character and one of itinerant character --- whose contrasting behavior under pressure is revealed by our measurements. A subset of electrons remains localized under strong Coulomb repulsion, sustaining \revised{an ordered moment} of about $1.4~\mu_{\mathrm{B}}$ that orders magnetically via exchange interactions and is robust against pressure. The remaining Fe~$3d$ electrons are weakly correlated and become increasingly itinerant with compression, in agreement with theoretical calculations performed on the \revised{related} compound \bfse~\cite{Craco2020}, enhancing the density of states at the Fermi level --- either by closing orbital-dependent gaps or by increasing hopping amplitudes as Fe--S bond lengths shorten. Once a sufficient density of itinerant states is achieved, superconductivity emerges.

This orbital-selective framework \revised{can reconcile} the persistence of magnetism with the emergence of superconductivity, without \revised{the need to invoke} phase separation~\cite{Li2012} or fine-tuned magnetic quantum criticality~\cite{Si2008,Zhou2013,Putzke2014}. \revised{Within this framework,} the nearly constant ordered moment, despite a dramatic reduction of the local moment, reflects the survival of the most strongly correlated orbitals. Notably, the structural transition near 1~GPa modifies exchange pathways and \revised{provides a natural explanation for} the reported anomaly in $T_{\mathrm{N}}$~\cite{Chi2016}.

A \revised{comparison} with \bfse\ is instructive. In the selenide, the ordered moment under pressure stabilizes at a higher value of about $2.8~\mu_{\mathrm{B}}$~\cite{Zheng2022}, \revised{which may indicate} a larger fraction of localized electrons. This \revised{would be} consistent with its lower superconducting $T_c$\revised{, as} fewer itinerant carriers \revised{would be} available for pairing. Conversely, the higher $T_c$ of \bfs\ \revised{could reflect} a larger proportion of itinerant electrons and a higher density of states at the Fermi level. \revised{This tentative correlation, based on only two compounds, should however be tested on a broader set of materials and by orbital-resolved probes.}

\revised{We note that none of our probes is orbital-resolved, and that the IAD of the $K\beta$ line is also sensitive to Fe--S covalency~\cite{Vanko2006a,Lafuerza2020}; the orbital-selective picture is therefore the most natural interpretation of our observations rather than a direct measurement. Neutron diffraction with a finer pressure mesh between 7.7 and 10~GPa, where $T_c$ is maximal, and orbital-resolved probes would further test this scenario.}

\revised{Our results thus indicate that} the coexistence of magnetism and superconductivity in \bfs\ \revised{does not result from} simple competition, but \revised{can be understood as arising} from the selective response of multiorbital electronic states to pressure. Taken together, our combined x-ray and neutron diffraction, x-ray emission spectroscopy, and electrical transport measurements map the structural, magnetic, and electronic evolution of \bfs\ across the insulator--metal transition and into the superconducting regime, revealing a pressure-driven decoupling between the robust stripe-type antiferromagnetic order and the strongly suppressed local Fe magnetic moment. This dichotomy \revised{indicates} that distinct subsets of Fe~$3d$ electrons respond differently to pressure --- itinerant carriers progressively delocalize and \revised{are expected to} give rise to superconductivity, whereas localized electrons continue to sustain long-range magnetic order. This \revised{makes} \bfs\ \revised{a} model system for orbital-selective physics in low-dimensional correlated materials, and more broadly a rare example in which superconductivity and magnetism coexist through the selective evolution of multiorbital electronic states.

\section*{Methods}

\subsection*{Sample synthesis}

Single crystals of \bfs\ were grown from the melt using the self-flux method~\cite{Amigo2021,Caron2011,Caron2012}. BaS (99.9\%), Fe (99.9\%), and S (99.999\%) powders were mixed in a BaS:Fe:S molar ratio of 1:2.05:3, pelletized, placed in a carbon crucible, and sealed in an evacuated quartz tube with a partial pressure of 300~mbar of Ar gas. The assembly was annealed at 1100\,$^\circ$C for 24~hours and subsequently cooled to 750\,$^\circ$C at a rate of 6\,$^\circ$C/h before cooling to room temperature. This yielded centimeter-sized pellets consisting of co-aligned millimeter-sized rod-shaped crystals along the $c$ direction. Phase purity was confirmed by powder x-ray diffraction, and the chemical composition was verified by energy-dispersive x-ray spectroscopy (SEM/EDX) and scanning transmission electron microscopy (STEM/EDX)~\cite{Oubaid2026}. All measurements reported here were performed on samples from the same synthesis batch.

\subsection*{Single-crystal x-ray diffraction under pressure}

Single-crystal x-ray diffraction under pressure was performed at the CRISTAL beamline (SOLEIL synchrotron) at $\lambda = 0.4175$~\AA, using a 2D detector and data processed with \textsc{CrysAlis\,Pro}. A DAC with 600~$\mu$m culets was used, with a sample chamber of 300~$\mu$m diameter and 50~$\mu$m thickness. Several small crystals (approximately 40~$\mu$m in size) were co-loaded with gold powder and a ruby chip. Helium was used as the pressure-transmitting medium to ensure quasi-hydrostatic conditions. Pressure was determined primarily by ruby fluorescence (relative accuracy $\sim$5\%)~\cite{Shen2020} and cross-checked against the gold equation of state. Low-temperature measurements were performed using a helium-circulation cryostat specially designed for DACs, allowing sample rotation at low temperature with in-situ pressure adjustment. Structural refinements were carried out using \textsc{JANA2020}. Refinement statistics and lattice parameters at all measured pressures are provided in the Supplementary Information.

\subsection*{Neutron diffraction under pressure}

Powder neutron diffraction under pressure was performed on the D1B (CRG) and D20 diffractometers at the Institut Laue-Langevin (ILL, Grenoble, France) with wavelengths of 2.52~\AA\ and 2.42~\AA, respectively. A Paris--Edinburgh pressure cell with a sample volume of approximately 50~mm$^3$ and an ethanol--methanol mixture as the pressure-transmitting medium was used. Lead (Pb) was co-loaded as a pressure marker; pressure was determined from the refined Pb lattice parameter combined with its equation of state~\cite{Strassle2014}. Crystal and magnetic structure refinements were carried out using the \textsc{FullProf} suite~\cite{Rodriguez-Carvajal1993}. Refinement results are provided in the Supplementary Information.

\subsection*{X-ray emission spectroscopy}

Fe~$K\beta$ x-ray emission spectroscopy was performed at the GALAXIES beamline~\cite{Rueff2015,Ablett2025} at the SOLEIL synchrotron. XES is a bulk-sensitive, element-selective technique that measures the instantaneous local spin state of the Fe~$3d$ shell through the exchange interaction between the Fe~$3p$ core hole and the $3d$ valence electrons, independently of long-range magnetic order. Pressure was applied using a membrane-driven DAC equipped with 1.7~mm-thick diamonds with 600~$\mu$m culets. Several \bfs\ single crystals were loaded in a 300~$\mu$m hole of a CuBe gasket, together with ruby chips for pressure calibration and an ethanol--methanol mixture as the pressure-transmitting medium. XES spectra were recorded with 10~keV incident x-rays, using a 1-m-radius spherically bent Ge(620) crystal analyser and an avalanche photodiode detector in the Rowland circle geometry, yielding an energy resolution of 1.2~eV FWHM at the Fe~$K\beta$ line ($\sim$7057~eV). The local spin state was quantified by the integrated absolute difference (IAD) method~\cite{Rueff2001}, using FeS$_2$ ($S = 0$) as a non-magnetic reference. Spectra were collected at 300~K as a function of pressure.

\subsection*{Electrical resistivity under pressure}

Temperature-dependent electrical resistivity was measured using the standard four-probe DC technique in a diamond-anvil cell. Daphne 7373 was used as the pressure-transmitting medium to ensure the highest possible degree of hydrostaticity; given the quasi-one-dimensional crystal structure of \bfs, non-hydrostaticity can suppress the superconducting volume fraction~\cite{Takahashi2015,yamauchi2015pressure}. Although Daphne 7373 undergoes a liquid-to-solid transition near 2.2~GPa at room temperature~\cite{Klotz2009}, it maintains a quasi-hydrostatic environment well into the GPa range. Pressure was calibrated in situ via ruby fluorescence at ambient temperature. Platinum wires (10~$\mu$m diameter) were used for electrical leads, fixed to the sample with silver epoxy. The Fermi-liquid quadratic coefficient $A$ was extracted via a least-squares fit of $\rho = \rho_0 + AT^2$ in the low-temperature range (2--40~K) at pressures exceeding the insulator-to-metal transition. Fermi-liquid fits are shown in the Supplementary Information.

\bibliography{Biblio}

\section*{Acknowledgements}

The authors gratefully acknowledge Professor C. Pasquier for fruitful discussions. The authors thank SOLEIL for synchrotron beam time on GALAXIES and CRISTAL beamlines (Proposal 20230425) and ILL for neutron beam time on D1B and D20 (Proposals 5-31-3014 and 5-31-2963; DOI:10.5291/ILL-DATA.5-31-3014 and DOI:10.5291/ILL-DATA.5-31-2963). This work was financially supported by the ANR COCOM 20-CE30-0029, the France 2030 programme ``ANR-11-IDEX-0003'' via Integrative Institute of Materials from Paris-Saclay University -- 2IM@UPSaclay, the Paris Ile-de-France Region in the framework of DIM MaTerRE (project DAC-VX), and the French Neutron Federation (2FDN). N.S.D. acknowledges support from the Croatian Science Foundation under project No. IP-2020-02-9666 and from the Cryogenic Centre at the Institute of Physics --- KaCIF (Grant No. KK.01.1.1.02.0012), co-financed by the Croatian Government and the European Union through the European Regional Development Fund --- Competitiveness and Cohesion Operational Programme.

\section*{Author contributions statement}

Y.O., P.Fo., C.R., and C.V.C. prepared and carried out the neutron diffraction experiment. Y.O., P.Fe., V.B., P.Fo. performed the x-ray diffraction experiment. V.B. and J.-P.R. performed the XES experiment. \revised{N.S.D.} performed the transport measurements. \revised{A.F. and D.B.} synthesized and characterized the samples. Y.O., P.Fe., P.Fo., and V.B. contributed to the analysis of the data. \revised{M.V. contributed to the design and follow-up of the project.} V.B. prepared the manuscript with input from all co-authors.

\section*{Additional information}

\textbf{Competing interests:} The authors declare no competing interests.

\textbf{Data availability:} \revised{The crystallographic data supporting this study (CIF files at all six pressures) have been deposited in the Cambridge Structural Database under deposition numbers CCDC~2543119--2543124 and are publicly available free of charge at \url{https://www.ccdc.cam.ac.uk/structures/}. The neutron diffraction data are available from the ILL data repository at \url{https://doi.org/10.5291/ILL-DATA.5-31-3014} and \url{https://doi.org/10.5291/ILL-DATA.5-31-2963}.} Other data supporting the findings of this study are available from the corresponding author upon reasonable request.

\end{document}

\begin{center}
{\Large \textbf{Supplementary Information for:}}\\[4mm]
{\Large \textbf{\revised{Coexistence of superconductivity and magnetic order in BaFe$_2$S$_3$: evidence for orbital-selective delocalization}}}\\[6mm]
Y. Oubaid, N. S. Dhami, P. Fertey, C. V. Colin, C. Ritter, A. Forget, D. Bounoua,\\
J.-P. Rueff, M. Verseils, P. Foury-Leylekian, V. Bal\'edent$^{*}$\\[2mm]
\textit{$^{*}$Corresponding author: victor.baledent@universite-paris-saclay.fr}
\end{center}

\vspace{6mm}

This Supplementary Information accompanies the Scientific Reports article ``\revised{Coexistence of superconductivity and magnetic order in BaFe$_2$S$_3$: evidence for orbital-selective delocalization}.'' It provides additional experimental details, refinement statistics, and supporting figures for the single-crystal x-ray diffraction, powder neutron diffraction, and electrical transport measurements under pressure discussed in the main text.

\section*{Supplementary Note 1: Single-crystal x-ray diffraction under pressure}

\begin{figure}[H]
\centering
\includegraphics[width=0.8\linewidth]{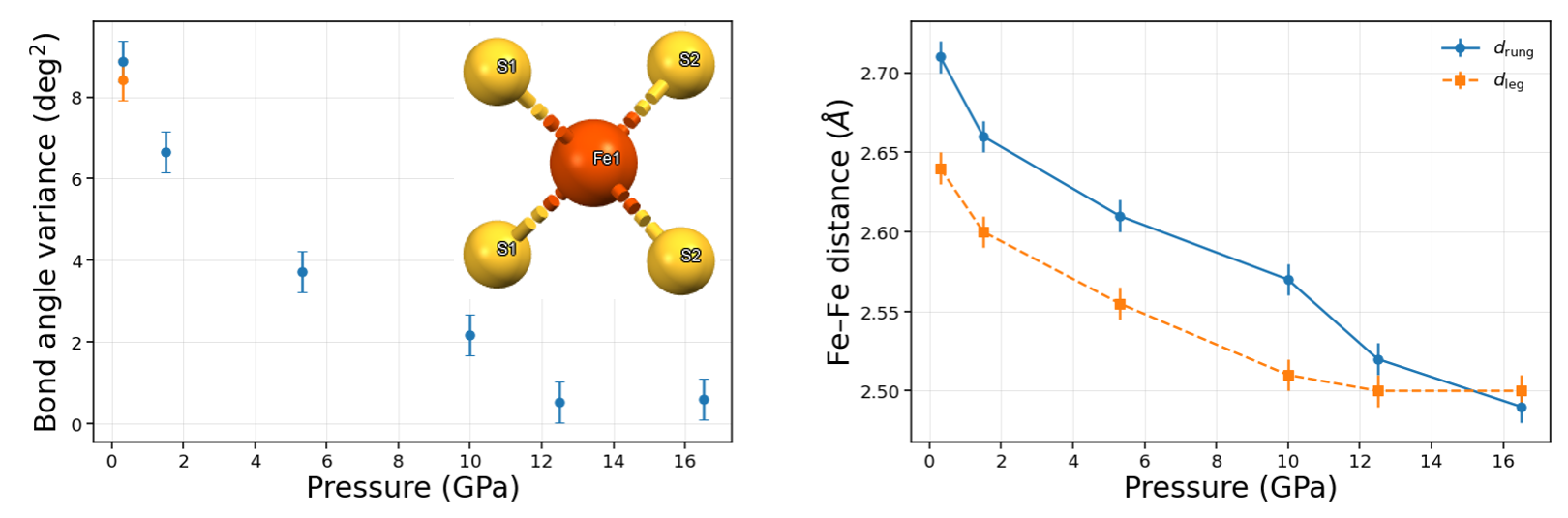}
\caption{(a) Bond--angle variance of the FeS$_4$ tetrahedra in \bfs\ as a function of pressure. Error bars represent one standard deviation propagated from the least-squares uncertainties of the bond angles. The orange symbol indicates the value for the symmetry-inequivalent Fe site in the $Pb2_1m$ symmetry. Inset: schematic FeS$_4$ tetrahedron. (b) Evolution of the Fe--Fe distances along the leg (orange symbols) and the rung (blue symbols) of the ladder.}
\label{Fig_tetrahedra}
\end{figure}

Single-crystal x-ray diffraction under pressure was performed at a wavelength of $\lambda = 0.4175$~\AA, using a 2D detector and data processed with \textsc{CrysAlis\,Pro}. A DAC with 600~$\mu$m culets was used, with a sample chamber of 300~$\mu$m diameter and 50~$\mu$m thickness. Several small crystals (approximately 40~$\mu$m in size) were co-loaded with gold powder and a ruby chip. Helium was used as the pressure-transmitting medium to ensure quasi-hydrostatic conditions. Pressure was determined primarily by ruby fluorescence (relative accuracy $\sim$5\%)~\cite{Shen2020} and cross-checked against the gold equation of state from the loaded Au powder. Low-temperature measurements were performed using a helium-circulation cryostat specially designed for DACs, allowing sample rotation at low temperature with in-situ pressure adjustment. Reciprocal space maps were reconstructed to identify symmetry changes and to monitor reflections governed by space-group extinction conditions. Structural refinements were carried out using \textsc{JANA2020} software. All CIF files are deposited at the Cambridge Structural Database (CSD), maintained by the Cambridge Crystallographic Data Centre (CCDC), under deposition \revised{numbers CCDC~2543119--2543124, and are publicly available free of charge at \url{https://www.ccdc.cam.ac.uk/structures/}.}

\begin{table}[H]
\centering
\caption{Combined table of crystal data and refinement details for different \textbf{pressures} at low temperature $15 \pm 3$~K, with the space group (Sp.\ Grp.), lattice parameters ($a$, $b$, $c$ and unit-cell volume $V$), number of reflections collected (Col.), independent (Ind.) and observed with intensity over $2\sigma$ ($I\ge 2\sigma(I)$), data-reduction metrics [$R_{\mathrm{int}}$, $(\sin\theta/\lambda)_{\max}$], and refinement statistics [$R(F^{2}>2\sigma(F^{2}))$, weighted $wR(F^{2})$, and goodness of fit $S$.}
\label{tab:pressure_params}
\vspace{4mm}
\begin{tabular}{|l|c|c|c|c|c|c|}
\hline
$P$ (GPa) & $0.3$ & $1.5$ & $5.5$& $10$ & $12.5$& $16.5$ \\ \hline
Sp.\ Grp & \textit{Pb2$_1$m} & \textit{Cmcm} & \textit{Cmcm} & \textit{Cmcm} & \textit{Cmcm} & \textit{Cmcm} \\ \hline
$a$ (\AA) & 8.787(13)& 8.6626(7)& 8.4867(12)& 8.4178(15)& 8.3784(7)& 8.3272(12)\\ \hline
$b$ (\AA) & 11.225(7)& 10.968(4)& 10.399(3)& 10.168(19)& 9.888(3)& 9.702(5)\\ \hline
$c$ (\AA) & 5.288(9)& 5.2116(5)& 5.1102(7)& 5.0366(11)& 5.0007(6)& 4.9993(7)\\ \hline
$V$ (\AA$^{3}$) & 521.58(3)& 495.16(19)& 450.99(16)& 431.09(13)& 414.29(16)& 403.9(2)\\ \hline
\multicolumn{7}{|c|}{\textbf{Number of Bragg peaks}} \\ \hline
Col. (measured) & 7779 & 1262 & 1556 & 2104 & 1491 & 1191 \\ \hline
Ind. (independent) & 731 & 173 & 146 & 144 & 143 & 134 \\ \hline
$I\ge 2\sigma(I)$ (observed) & 682 & 171 & 144 & 143 & 142 & 127 \\ \hline
\multicolumn{7}{|c|}{\textbf{Data reduction}} \\ \hline
$R_{\mathrm{int}}$ & 0.126 & 0.077 & 0.131 & 0.126 & 0.098 & 0.061 \\ \hline
$(\sin\theta/\lambda)_{\max}$ (\AA$^{-1}$) & 0.627 & 0.622 & 0.615 & 0.617 & 0.623 & 0.641 \\ \hline
\multicolumn{7}{|c|}{\textbf{Refinement}} \\ \hline
$R\!\left[F^{2}>2\sigma(F^{2})\right]$ & 0.040 & 0.035 & 0.074 & 0.038 & 0.052 & 0.057 \\ \hline
$wR(F^{2})$ & 0.116 & 0.087 & 0.195 & 0.105 & 0.154 & 0.145 \\ \hline
$S$ & 1.03 & 1.05 & 1.01 & 1.06 & 1.01 & 1.01 \\\hline
\end{tabular}
\end{table}

\section*{Supplementary Note 2: Neutron diffraction under pressure}

We performed powder neutron diffraction under pressure, using powders prepared by grinding single crystals of \bfs. Measurements were performed at the D1B and D20 diffractometers at the Institut Laue--Langevin, with an incident wavelength of $2.52$~\AA\ for D1B and $2.42$~\AA\ for D20. A Paris--Edinburgh press (sample volume $\sim 50~\mathrm{mm}^3$) with an ethanol--methanol mixture as the pressure medium was employed, reaching a maximum pressure of $10~\mathrm{GPa}$. Lead was included as an internal standard to determine the pressure from its reference diffraction peaks and its standard equation of state (EOS). The error bars for each pressure point are: $2\pm0.3$~GPa, $5.5\pm0.3$~GPa, $7.7\pm0.2$~GPa, and $9.9\pm0.2$~GPa. Magnetic-structure refinements were performed using the \textsc{FullProf} Suite~\cite{Rodriguez-Carvajal1993}.

\begin{figure}[H]
\centering
% Width reduced so that the figure and its full caption fit on one page (caption was truncated)
\includegraphics[width=0.55\linewidth]{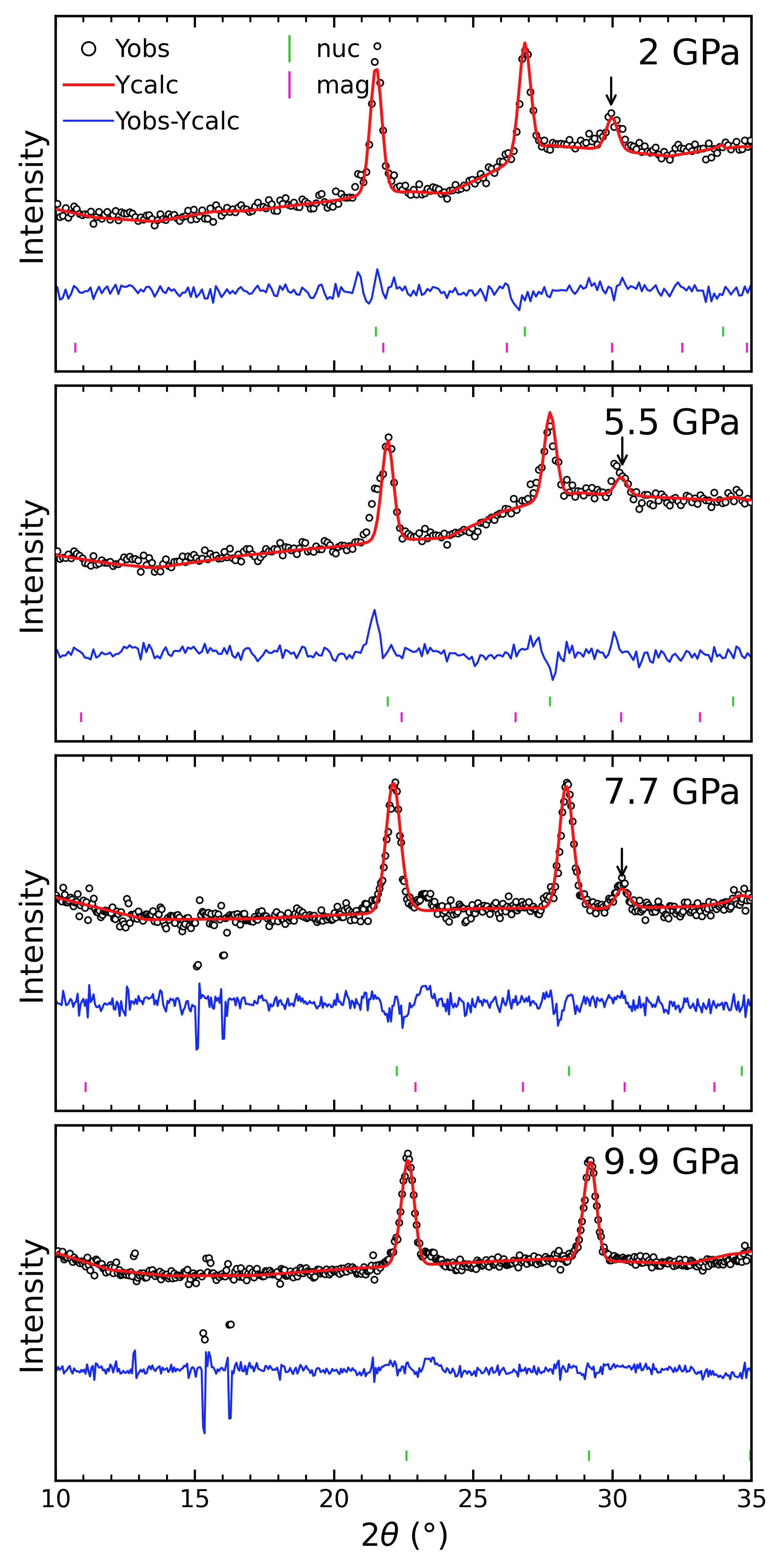}
\caption{Rietveld refinement of the low-angle neutron diffraction patterns of \bfs\ at 10~K (20~K for the 2 and 5.5~GPa data) under pressure. For each pressure, the observed data are shown as open circles, the calculated pattern as a solid red line, and the difference curve $Y_{\mathrm{obs}}-Y_{\mathrm{calc}}$ as a blue line. Vertical tick marks denote the positions of the nuclear (green) and magnetic (magenta) Bragg reflections. Only the low-angle region is shown in order to emphasize the pressure dependence of the magnetic contribution. The $2\theta$ values were converted to a common wavelength for a direct comparison between the powder patterns collected under different experimental conditions (D1B and D20 instruments). The vertical arrow indicates the magnetic Bragg peak $\left(\tfrac{1}{2}\ \tfrac{1}{2}\ 1\right)$.}
\label{Fig_PND}
\end{figure}

Supplementary Fig.~\ref{Fig_PND} shows the powder neutron patterns under different pressures. All patterns were collected at low temperature (20~K for 2~GPa and 10~K for the rest) after the pressurization and cooling processes. We performed powder neutron diffraction under pressure ($p$): low-$p$ measurements ($p = 2$ and $5.5$~GPa) at the \textsc{D1B} instrument, and high-$p$ measurements ($7.7$ and $9.9$~GPa) at the \textsc{D20} instrument. The magnetic Bragg reflection at $\left(\tfrac{1}{2}\ \tfrac{1}{2}\ 1\right)$ (the fingerprint of the stripe-like order) persists with nearly constant intensity up to $7.7$~GPa and then disappears at $9.9$~GPa.

Our refinement strategy consisted of using Rietveld refinements~\cite{Rietveld1969} for the nuclear phase up to $120^\circ$ in $2\theta$, then fixing the atomic positions and refining the additional magnetic phase using the same stripe model as at ambient pressure~\cite{Chi2016,Takahashi2015,Zheng2018} within the restricted range of $2\theta_{\max} = 35^\circ$, as the magnetic information comes from the magnetic Bragg peak $\left(\tfrac{1}{2}\ \tfrac{1}{2}\ 1\right)$ around $2\theta = 30^\circ$ (indicated by an arrow in Fig.~\ref{Fig_PND}), yielding an ordered moment of $m = 1.4 (\pm 0.2)\,\mu_B$ (1.4~$\mu_B$ for 2 and 5.5~GPa and 1.3~$\mu_B$ for 7.7~GPa) at all investigated pressures, statistically indistinguishable from the ambient-pressure value. After the refinement of each pattern, we obtained the lattice parameters and magnetic moment ($m$) at different pressures (Supplementary Table~\ref{tab:supp_pressure_refinement}).

\begin{table}[H]
\centering
\caption{Lattice parameters of \bfs, the magnetic ordered moment, and magnetic agreement factor extracted from the FullProf refinement files.}
\vspace{2mm}
\begin{tabular}{cccccc}
\hline\hline
\noalign{\vskip 0.6mm}
Pressure (GPa) & $a$ (\AA) & $b$ (\AA) & $c$ (\AA) & $M_{Fe}$ ($\mu_B$) & $R(\mathrm{mag.})$ (\%) \\
\hline
2(\textpm 0.3)& 8.652(36) & 10.886(24) & 5.238(40) & 1.4(2) & 46.3 \\
5.5(\textpm 0.3)& 8.564(43) & 10.537(28) & 5.190(28) & 1.4(2) & 9.3 \\
7.7(\textpm 0.2)& 8.4609(2) & 10.2587(4) & 5.1595(1) & 1.3(2) & 11.44 \\
9.9(\textpm 0.2)& 8.3903(2) & 10.0105(3) & 5.1176(1) & --- & --- \\
\hline\hline
\end{tabular}
\label{tab:supp_pressure_refinement}
\end{table}

\section*{Supplementary Note 3: Resistivity under pressure}

\begin{figure}[H]
\centering
\includegraphics[width=0.8\linewidth]{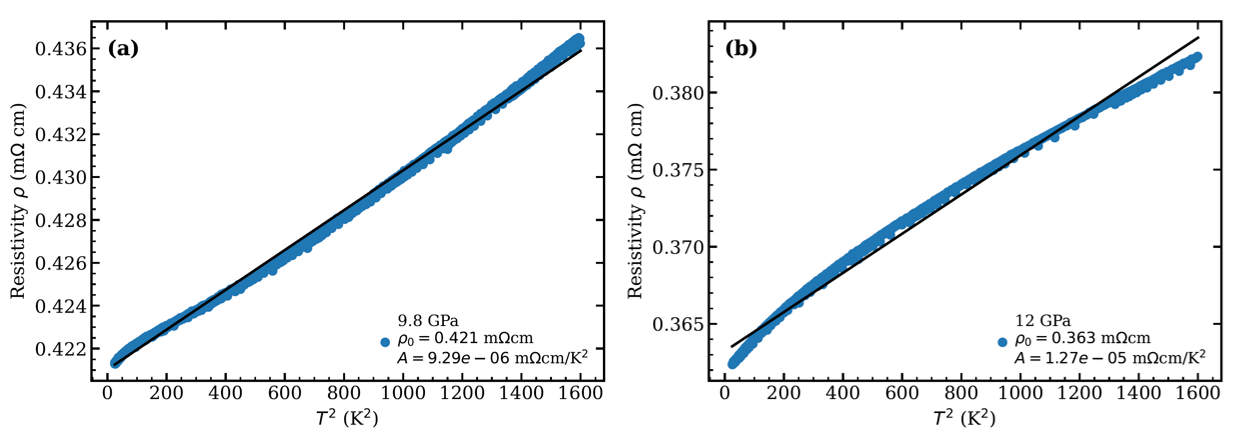}
\caption{Temperature dependence of the electrical resistivity of \bfs\ at 9.8~GPa (a) and 12~GPa (b) (blue circles), together with a quadratic fit ($\rho = \rho_0 + AT^2$) indicative of Fermi-liquid behavior (black curve).}
\label{Fig_FL}
\end{figure}

\bibliographystyle{unsrt}
\bibliography{Biblio}